%% file: Dhillon_WCL2017-0337.tex
\documentclass[final]{IEEEtran}
\pdfoutput=1
\usepackage{amsthm,amssymb,graphicx,multirow,amsmath,color,amsfonts}
\usepackage[update,prepend]{epstopdf}
\usepackage[noadjust]{cite}
\usepackage[latin1]{inputenc}
\usepackage{tikz}
\usepackage{bbm} 
\usepackage{pdfpages}
\usepackage{tabulary}
\usepackage{multirow}
\usepackage{comment}

\include{notation}

\allowdisplaybreaks 

\begin{document}
\graphicspath{{./Figures/}}
\title{
Tight Lower Bounds on the Contact Distance Distribution in Poisson Hole Process}
\author{
Mustafa A. Kishk and Harpreet S. Dhillon
\thanks{M. A. Kishk and H. S. Dhillon are with Wireless@VT, Department of ECE, Virginia Tech, Blacksburg, VA. Email: \{mkishk, hdhillon\}@vt.edu. The support of the U.S. National Science Foundation (Grants CCF-1464293 and CNS-1617896) is gratefully acknowledged. \hfill Last updated: \today.} 
\vspace{-10mm}
}

\maketitle
\begin{abstract}
In this letter, we derive new lower bounds on the cumulative distribution function (CDF) of the contact distance in the Poisson Hole Process (PHP) for two cases: (i) reference point is selected uniformly at random from $\nbbR^2$ independently of the PHP, and (ii) reference point is located at the center of a hole selected uniformly at random from the PHP. While one can derive upper bounds on the CDF of contact distance by simply ignoring the effect of holes, deriving lower bounds is known to be relatively more challenging. As a part of our proof, we introduce a tractable way of {\em bounding} the effect of all the holes in a PHP, which can be used to study other properties of a PHP as well.
\end{abstract}
\vspace{-3mm}
\section{Introduction} \label{sec:intro}
Owing to its ability to provide useful system-level performance insights, stochastic geometry has emerged as an attractive tool for the modeling and analysis of a variety of wireless networks~\cite{haenggi2012stochastic}. While Poisson Point Process (PPP) is often the default choice due to its simplicity and tractability, its inability to model inter-point interactions makes it an unrealistic choice for modeling scenarios in which locations of active nodes exhibit spatial separation due to interference management. One such general class of scenarios results from the creation of exclusion zones around wireless nodes/links to control aggregate interference received by them from the rest of the network. A more appropriate model for such scenarios is a PHP~\cite{6155562,7557010}, which is the main focus of this letter.

While one can introduce spatial separation between points in a point process in countless ways, PHP is perhaps one of the most tractable amongst them. It is formed by {\em carving out} holes from a baseline PPP, where the centers of the holes are assumed to follow an independent PPP. The remaining points of the baseline PPP are said to form a PHP. This model has found numerous applications in wireless networks. It was used in~\cite{6155562} to model a cognitive radio network, where the hole centers model the locations of the primary receivers and the PHP models the locations of the secondary transmitters. More recently, a very similar setup was used to model underlay device-to-device (D2D) communication in cellular networks, where protection zones are created around cellular links to save them from excessive interference from the D2D links~\cite{sun2014d2d,elsawy2014analytical,AfsDhiC2015a}. Similarly, PHP has also found application in the modeling of inter-tier dependence in a heterogeneous cellular network, where the hole centers represent macrocell base stations (BSs) and the PHP models the (active) small cell locations~\cite{7110505,6706236}. For a more detailed literature survey, interested readers are advised to refer to~\cite{7557010}. For completeness, please note that PHP is also sometimes referred to as the {\em Hole-$1$ process}~\cite{ganti2006regularity}. 

Even though PHP is generated from two independent homogeneous PPPs, its analysis is known to be significantly more challenging compared to a homogeneous PPP. In fact, until recently, the state-of-the-art approach to their analysis was to approximate them using homogeneous PPP whose density is matched to that of the PHP. One can also {\em bound} the performance of a PHP network by ignoring all the holes and approximating the PHP by its baseline PPP. More accurate bounding techniques were recently developed in \cite{7557010} that resulted in tight upper and lower bounds on the Laplace transform of interference in a PHP network.

Despite these recent efforts, we still lack complete characterization of several basic properties of a PHP. One of the most important amongst them is the contact distance distribution. In the literature, contact distance distribution is usually approximated by Weibull distribution whose parameters are determined using curve fitting~\cite{7110505}. While this approximation is usually tight, the use of curve fitting curtails the generality of this approach. In particular, since the parameters of Weibull distribution depend upon the system setup, we need to perform the curve fitting step every time the system parameters are changed. Second, somewhat less used approach, is to bound the CDF of contact distance from above by ignoring all the holes. The bound can be tightened slightly with some loss in tractability by incorporating the effect of the nearest hole to the reference point, as done in~\cite{yazd_conf}. In this letter, we derive the first known lower bounds on the CDF, which along with these upper bounds and approximations provide almost complete characterization of the contact distance distribution in a PHP. 

\textit{Contributions}. We derive lower bounds on the CDF of the contact distance in a PHP for two cases. In the first case, the reference point is chosen uniformly at random from $\mathbb{R}^2$ independently of the PHP, which is usually how the contact distance is classically defined. In the second case, the reference point is chosen to be the center of a hole selected uniformly at random from the PHP. This allows us to study the distance between the node which is being protected from the interference (located at the center of the hole) and the strongest interferer (closest point of the PHP from this node). In order to derive the lower bounds on CDF, we carefully {\em bound} the effect of carving out holes using simple geometric tricks that lend tractability to the analysis. A closed-form lower bound for the first case is also derived. The tightness of all the bounds is verified by comparing them with simulation results. 

\vspace{-2mm}
\section{Contact Distance Distributions}
Before deriving the distributions of the contact distance for the two reference points, we first formally define the PHP. 
\vspace{-2mm}
\subsection{Poisson Hole Process}
The PHP is constructed using two independent PPPs $\Phi_{1}\equiv\{y\}\subset\mathbb{R}^2$ and $\Phi_{2}\equiv\{x\}\subset\mathbb{R}^2$ with densities $\lambda_1$ and $\lambda_2$, respectively. The first PPP $\Phi_{1}$ represents the locations of the hole centers, while the second PPP $\Phi_2$ represents the baseline process from which the holes are carved out. The points retained in $\Phi_2$ after carving out the holes form the PHP $\Psi$, which can be mathematically defined as
\begin{align}
\Psi&=\{x\in\Phi_2:x\notin\bigcup_{y\in\Phi_1}\mathcal{B}(y,D)\},
\end{align}
where $D$ is the radius of the holes, and $\mathcal{B}(y,D)$ is a circle of radius $D$ centered at $y$. Using this notation, we now derive the lower bounds on the CDFs of the contact distance for two different cases. 
Due to the stationarity of this setup, we will place the reference point at the origin in both the cases.

\subsection{Reference Point is Chosen Uniformly at Random from $\mathbb{R}^2$}

In this case, we assume that the reference point is chosen uniformly at random from $\mathbb{R}^2$ independently of the PHP $\Psi$. Contact distance $R_1$ for this case is the distance between this reference point and its nearest point of $\Psi$. 
For this case, the CDF of the contact distance is defined next.
\begin{definition}[Contact distance distribution] \label{def:1}
The CDF of the contact distance when the reference point is chosen uniformly at random from $\mathbb{R}^2$ independently of $\Psi$ is
\begin{align}
F_{R_1}(r)=\mathbb{P}\left(R_1 < r\right)=\mathbb{P}\left(\mathcal{N}_{\Psi}\left(\mathcal{B}(o,r)\right) > 0\right),
\label{eq:cont_dist}
\end{align}
where $R_1$ is the contact distance, and $\mathcal{N}_{\Psi}\left(\mathcal{B}(o,r)\right)$ is the number of points of the PHP $\Psi$ inside a circle of radius $r$ centered at the origin.
\end{definition}
A lower bound on the CDF of $R_1$ is derived next.
\begin{theorem}[Lower bound on $F_{R_1}(r)$]\label{thm:random_point}
A lower bound on the CDF of contact distance $R_1$ is
\begin{align}
F_{R_{1}}(r)\geq 1-\exp\left(-\lambda_2\pi r^2\right)\exp\left(-2\pi\lambda_1\mathcal{G}_1(r)\right),
\end{align}
where $\mathcal{G}_1(r)=\left(1-\exp\left(\lambda_2\pi \min\{r,D\}^2\right)\right)\frac{(D-r)^2}{2}+\int_{|r-D|}^{D+r}\left(1-\exp(\lambda_2\mathcal{H}_1(r,r_y))\right)r_y{\rm d}r_y$,  
$|r-D|$ is the absolute value of $r-D$, $\mathcal{H}_1(r,r_y)=\left((r+D)^2-(r_y)^2\right)\theta(r_y)$, and $\theta(r_y)=\sin^{-1}\left(\frac{D}{D+r_y}\right)$.
\end{theorem}
\begin{IEEEproof}
See Appendix~\ref{app:random_point}.
\end{IEEEproof}
As discussed in detail in Appendix~\ref{app:random_point}, the above result is derived using the simple idea of excluding the area covered by the holes from the circle $\mathcal{B}(o,r)$. In order to maintain tractability, we neglect the possible overlaps between holes. As a result, we end up excluding the overlap regions multiple times, which provides a lower bound on the CDF. More details about the derivation can be found in Appendix~\ref{app:random_point}. 

While the lower bound presented in Theorem~\ref{thm:random_point} is fairly straightforward to compute, it is not in closed form since the expression for $\mathcal{G}_1(r)$ contains an integral term. The main challenge in simplifying the integral term in $\mathcal{G}_1(r)$ is the presence of $\sin^{-1}\left(\frac{D}{D+r_y}\right)$ term in the integrand. That being said, we can bound this expression by partitioning the integration interval into $N$ sub-intervals. For each of these sub-intervals, we can use the lower limit in the integral to get an upper bound on the $\sin^{-1}\left(\frac{D}{D+r_y}\right)$ term. This eventually leads to a lower bound on the result in Theorem~\ref{thm:random_point}. Obviously, as the value of $N$ increases, this lower bound will converge to the result provided in Theorem~\ref{thm:random_point}. As an example, we provide in the next Corollary the closed form bound for $N=1$ obtained using this procedure. It is straightforward to get the corresponding expression for any given value of $N$.%
\begin{cor} \label{cor:random_point}
A closed-form lower bound on the contact distance distribution of PHP is $F_{R_{1}}(r)\geq$
\begin{align}
\left\{
	\begin{array}{ll}
		 1-\exp\left(-\lambda_2\pi r^2\right)\exp\left(-2\pi\lambda_1\mathcal{G}_2(r)\right)&, {r}\leq D \\
		1-\exp\left(-\lambda_2\pi D^2\right)\exp\left(-2\pi\lambda_1\mathcal{G}_3(r)\right)&, {r} > D
	\end{array}.
	\right.
\end{align}
where $\mathcal{G}_2(r)=\mathcal{F}(\theta_1)$, $\mathcal{G}_3(r)=\mathcal{F}(\theta_2)$, $\theta_1=\sin^{-1}\left(\frac{D}{2D-r}\right)$, $\theta_2=\sin^{-1}\left(\frac{D}{r}\right)$, and $\mathcal{F}(\theta)=\left(1-\exp\left(\lambda_2\pi \min\{r,D\}^2\right)\right)\frac{(D-r)^2}{2}-\frac{\exp\left(4\theta\lambda_2 Dr\right)-1}{2\lambda_2\theta}+2rD$.
\end{cor}
\vspace{-6mm}
\subsection{Reference Point is one of the Hole Centers}
As explained already, hole centers in a wireless network often correspond to the nodes that are being protected from excessive interference. It is therefore important to study the statistics of the distance between a hole center and the closest point of $\Psi$ (closest possible interferer), which corresponds to the contact distance from a reference point placed at the center of a hole chosen uniformly at random from a PHP. We denote this distance by $R_2$ and its CDF is formally defined next. 
\begin{definition}[Distribution of the contact distance from a hole center] \label{def:2}
The CDF of the contact distance when the reference point belongs to $\Phi_1$ is
\begin{align}
F_{R_{2}}(r)=\mathbb{P}\left(R_{2} < r\right)=\mathbb{P}\left(\mathcal{N}_{\Psi}\left(\mathcal{B}(o,r)\right) > 0\Big|o\in\Phi_1\right),
\label{eq:cont_dist}
\end{align}
where $R_{2}$ is the contact distance, and $\mathcal{N}_{\Psi}\left(\mathcal{B}(o,r)\right)$ is the number of points of the PHP $\Psi$ that fall inside a circle of radius $r$ centered at the origin.
\end{definition}
By construction, the minimum distance between the hole center and its nearest PHP point is $D$. Using this fact and following the same approach used in Theorem~\ref{thm:random_point}, we derive a lower bound on the contact distance distribution for this case in the following Theorem.
\begin{theorem}[Lower bound on $F_{R_2}(r)$]\label{thm:hole_center}
A lower bound on the CDF of the contact distance $R_2$ is: $F_{R_{2}}(r)$
\begin{align}
\geq	
\begin{array}{ll}
		1-\exp\left(-\lambda_2\pi (r^2-D^2)\right)\exp\left(-2\pi\lambda_1\mathcal{G}_4(r)\right) &, {r} > D
	\end{array}.
\end{align}
where $\mathcal{G}_4(r)=\int_0^{r+D}\left(1-\exp(\lambda_2\mathcal{H}_2(r,r_y))\right)r_y{\rm d}r_y,$ and\\$\mathcal{H}_2(r,r_y)=\left(\min\{r+D,r_y+2D\}^2-\max\{r_y,2D\}^2\right)\theta(r_y)$. 
\end{theorem}
\begin{IEEEproof}
See Appendix~\ref{app:hole_center}.
\end{IEEEproof}
As was the case in Theorem~\ref{thm:random_point}, it is fairly straightforward to compute the lower bound presented in the above Theorem. The accuracy of both these result as well as the closed-form lower bound for the previous case is investigated next.


\subsection{Numerical Results}
The tightness of the lower bounds on the CDFs of both $R_1$ and $R_2$ is verified by comparing them with simulations in Figs.~\ref{fig:1} and~\ref{fig:2}. We consider $\lambda_1=10$ ${\rm km^{-2}}$ and several combinations of $\lambda_2$ and $D$, which are all indicated in the plots. The closed-form lower bound on the CDF of $R_1$ is also plotted for completeness. We chose $N=8$, which was sufficient in this case to converge to the result of Theorem~\ref{thm:random_point}. In order to provide a complete picture, we also include an upper bound on the CDF, which is computed by {\em bounding} the PHP by its baseline PPP $\Phi_2$, and an {\em approximation} which is computed by approximating the PHP by a PPP with equivalent density $\tilde{\lambda}=\lambda_2e^{-\lambda_1\pi D^2}$. In both these PPP-based cases, the contact distance distribution is easily computed using the null probability of the PPP. In both the figures, we notice that new lower bounds along with the PPP-based upper bound and heuristic approximation collectively provide a sharp characterization of the contact distance distribution.

\vspace{-3mm}
\section{Conclusion and Future Work}
In this letter, we derived tight lower bounds on the CDF of the contact distance in a PHP for two different choices of the reference point.  The main technical contribution is a new bounding technique using which we carefully handled the effect of all the holes of a PHP while maintaining tractability. The tightness of the new bounds across different scenarios is verified numerically. The proposed approach and new results can be readily used to derive tight bounds for key performance metrics of interest, such as receiver power, coverage probability, and throughput, in wireless networks modeled as a PHP.%

\vspace{-3mm}
\appendix
In all our derivations we will refer to $\mathcal{H}_1(r,r_y)$ and $\mathcal{H}_2(r,r_y)$ which are defined in Theorems~\ref{thm:random_point} and~\ref{thm:hole_center}, respectively. In addition, the reference point is assumed at the origin $o$, while $y$ represents the location of an arbitrary hole center.
\vspace{-3mm}
\subsection{Proof of Theorem~\ref{thm:random_point}}\label{app:random_point}
\begin{figure}
\centering
\includegraphics[width=0.5\columnwidth]{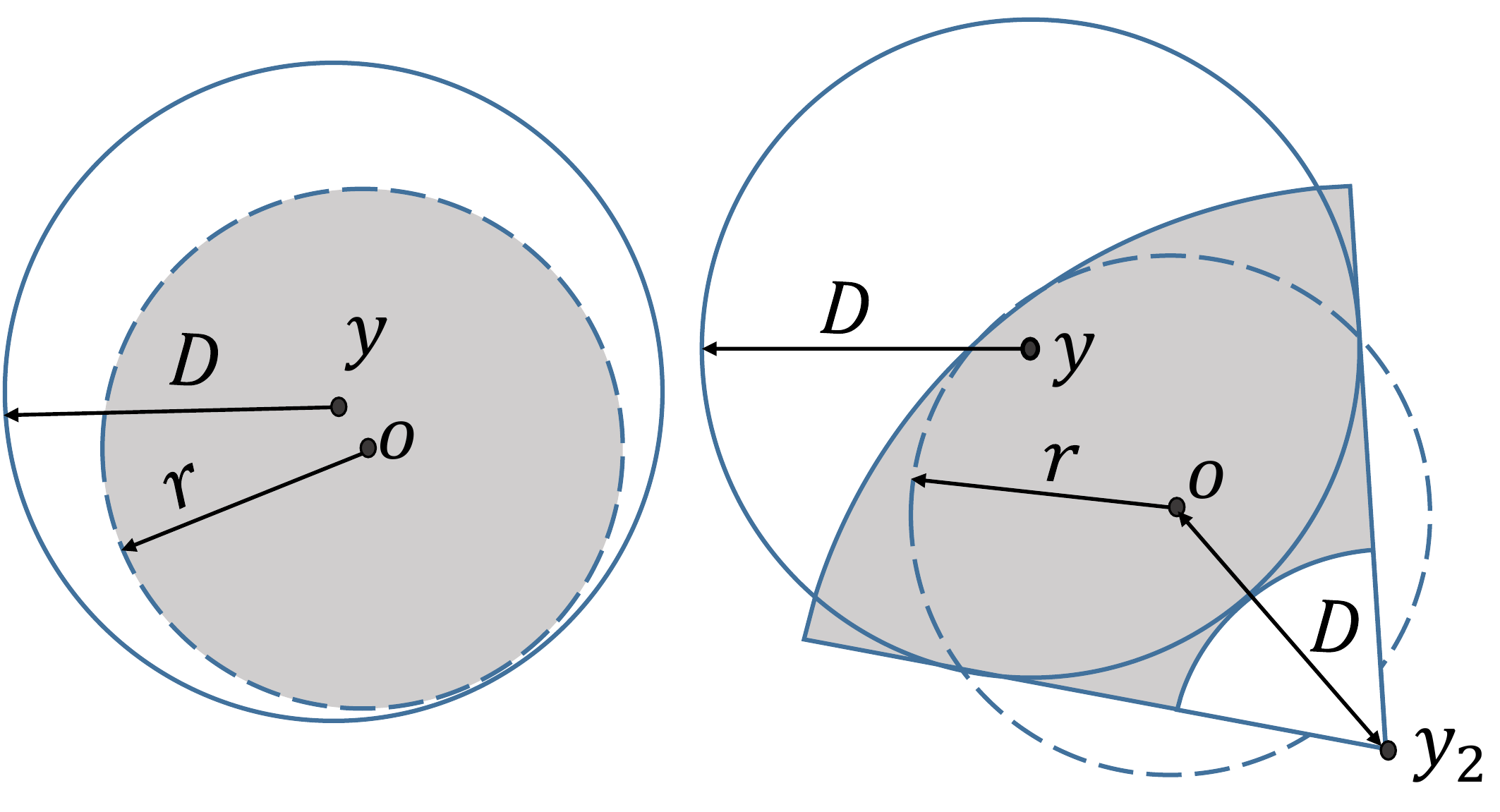}
\caption{The areas covered by $\tilde{A}(r)$ when the reference point is chosen uniformly at random from $\mathbb{R}^2$ independently of $\Psi$ ($r\leq D$).}
\label{fig:3}
\vspace{-3mm}
\end{figure}

To derive a lower bound on $F_{R_1}(r)$, we need an upper bound on the CCDF $\bar{F}_{R_1}(r)=1-F_{R_1}(r)=\mathbb{P}(R_1>r)$. The exact expression of the CCDF is
\begin{align}
\bar{F}_{R_1}(r)&=\mathbb{E}_{\Phi_1}\mathbb{E}_{\Phi_2}\left[\mathbbm{1}\left(\mathcal{N}_{\Phi_2}\left(\mathcal{B}(o,r)\setminus\mathcal{A}(r)\right)=0\bigg|\Phi_1,\Phi_2\right)\right]\nonumber\\
&=\mathbb{E}_{\Phi_1}\left[\mathbb{P}\left(\mathcal{N}_{\Phi_2}\left(\mathcal{B}(o,r)\setminus\mathcal{A}(r)\right)=0\bigg|\Phi_1\right)\right],
\end{align}
where $\mathcal{N}_{\Phi_2}(\Xi)$ is the number of points of $\Phi_2$ in the area covered by any {\em generic} region $\Xi \subset \mathbb{R}^2$, $\mathcal{A}(r)=\mathcal{B}(o,r)\cap\mathcal{A}_1$, and $\mathcal{A}_1=\bigcup_{y\in\Phi_1}\mathcal{B}(y,D)$. The region $\mathcal{A}_1$ represents the whole area covered by the holes, and $\mathcal{A}(r)$ is the portion of this area enclosed within the circle centered at the origin with radius $r$. Modeling $\mathcal{A}_1$ is the main challenge in this analysis. Since we are interested in getting an upper bound on $\bar{F}_{R_1}(r)$, we will use an upper bound on the area of the region $\mathcal{A}_1$ which is $|\tilde{\mathcal{A}}_1|=\sum_{y\in\Phi_1}|\mathcal{B}(y,D)|$, where $|\Xi|$ is the area of the region $\Xi$. This upper bound on $\mathcal{A}_1$ overestimates the area covered by the holes since it accounts for the overlaps between holes multiple times. Hence, it overestimates the CCDF $\bar{F}_{R_1}(r)$ leading to an upper bound. Defining an upper bound on $\mathcal{A}(r)$ as $\tilde{\mathcal{A}}(r)=\mathcal{B}(o,r)\cap\tilde{\mathcal{A}}_1$, we have
\begin{align}
\bar{F}_{R_1}(r)&\leq \mathbb{E}_{\Phi_1}\left[\mathbb{P}\left(\mathcal{N}_{\Phi_2}\left(\mathcal{B}(o,r)\setminus\tilde{\mathcal{A}}(r)\right)=0\bigg|\Phi_1\right)\right]\nonumber\\
&\stackrel{(a)}{=} \mathbb{E}_{\Phi_1}\left[\exp\left(-\lambda_2\left(\pi r^2-|\tilde{\mathcal{A}}(r)|\right)^{+}\right)\right]\nonumber\\
&\stackrel{(b)}{\leq} \mathbb{E}_{\Phi_1}\left[\exp\left(-\lambda_2\left(\pi r^2-|\tilde{\mathcal{A}}(r)|\right)\right)\right]\nonumber\\
&= \exp\left(-\lambda_2\pi r^2\right)\mathbb{E}_{\Phi_1}\left[\exp\left(\lambda_2|\tilde{\mathcal{A}}(r)|\right)\right],
\end{align}
\begin{figure}
\centering
\includegraphics[width=0.6\columnwidth]{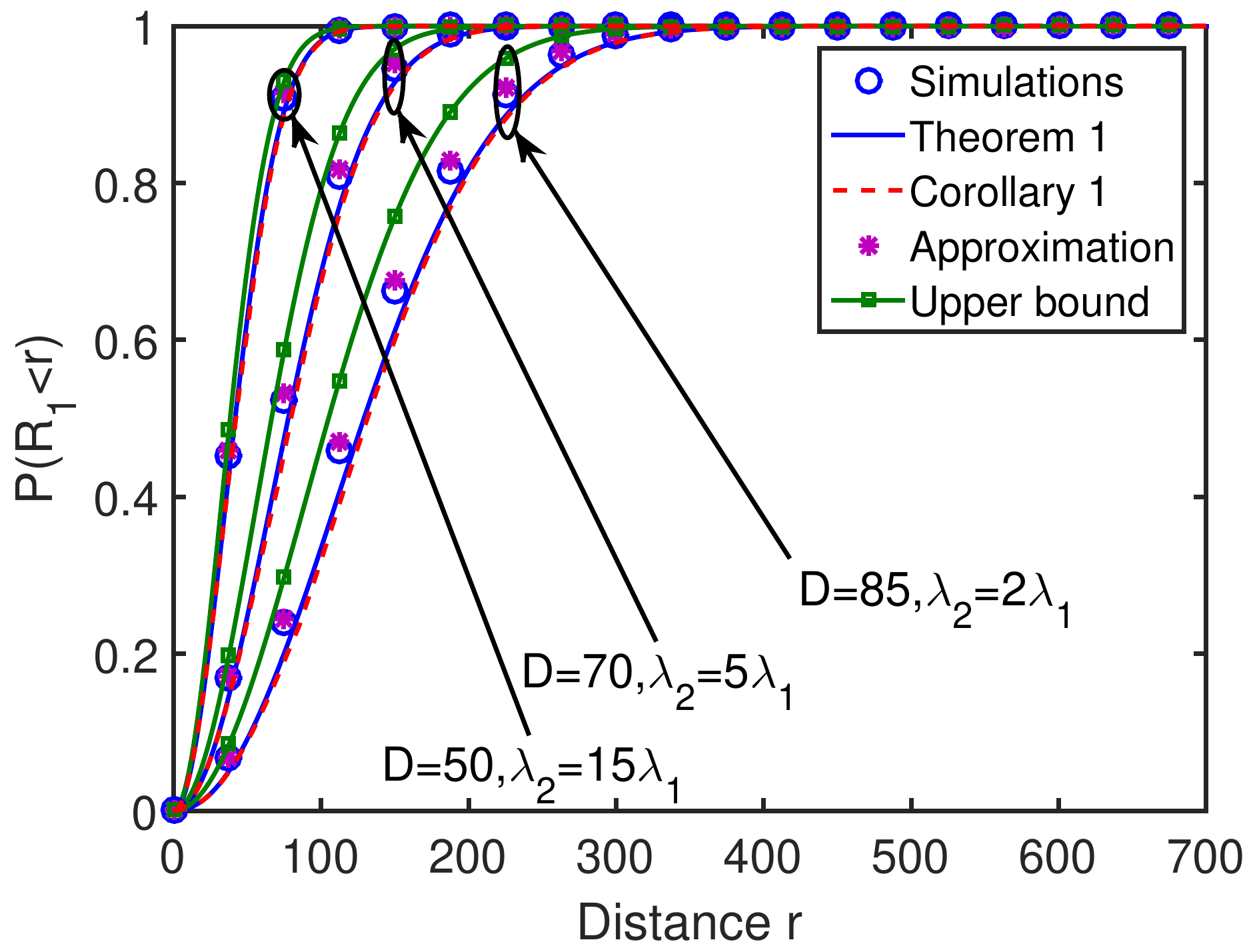}
\caption{The CDF of the contact distance $R_1$.}
\label{fig:1}
\end{figure}
\begin{figure}
\centering
\includegraphics[width=0.6\columnwidth]{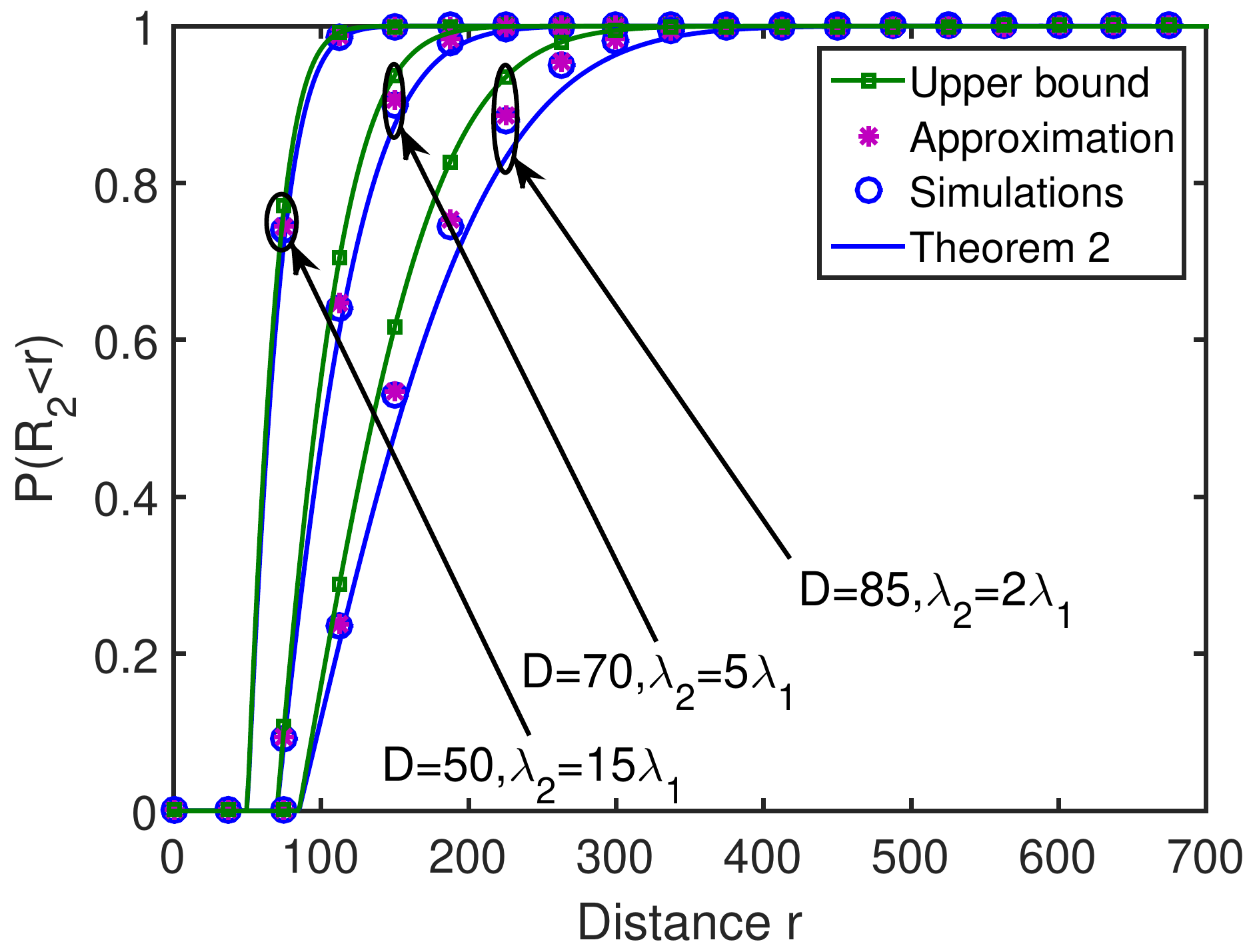}
\caption{The CDF of the contact distance $R_2$.}
\label{fig:2}
\vspace{-6mm}
\end{figure}
where step (a) results from the null probability of the homogeneous PPP $\Phi_2$ ($x^{+}=\max\{0,x\}$ in this step). Step $(b)$ follows from that fact that $x^{+}\geq x$. The main challenge in the rest of this derivation is to accurately determine $|\tilde{\mathcal{A}}(r)|$ for different values of $r$ while maintaining tractability. For the case of $r<D$, the area covered by $\tilde{\mathcal{A}}(r)$ is represented by two types of holes. The first type is when $\|y\|<D-r$. In that case, as shown in Fig.~\ref{fig:3} (left), the circle $\mathcal{B}(o,r)$ is completely enclosed inside the hole. The second type is when $D-r<\|y\|<D+r$. In that case, as shown in Fig.~\ref{fig:3} (right), we need to model the intersection between the hole and the circle $\mathcal{B}(o,r)$. To facilitate that, we assume a virtual point $y_2$ at distance $D$ from the origin. From this point, we draw two lines tangent to the hole that encloses this intersection. This leads to the shaded area shown in Fig.~\ref{fig:3} (right), which is $\mathcal{H}_1(r,\|y\|)$. Although the shaded area is larger than the required intersection, it gives much more tractable expressions. Hence, when $r<D$, the CCDF is upper bounded as follows
\begin{align}
\label{10}
&\bar{F}_{R_1}(r)\leq \exp\left(-\lambda_2\pi r^2\right)\mathbb{E}_{\Phi_1}\Bigg[\exp\Bigg(\lambda_2 \Bigg(  \sum_{y\in\Phi_1\cap\mathcal{B}(o,D-r)}\pi r^2+\nonumber\\
&\sum_{y\in\Phi_1\cap\mathcal{B}(o,D-r)^{c}\cap\mathcal{B}(o,r+D)}\mathcal{H}_1(r,\|y\|) \Bigg)  \Bigg)\Bigg]\nonumber\\
&= \exp\left(-\lambda_2\pi r^2\right)\mathbb{E}_{\Phi_1}\Bigg[\prod_{{y\in\Phi_1\cap\mathcal{B}(o,D-r)}}\exp(\lambda_2\pi r^2)\times\nonumber\\
&\prod_{y\in\Phi_1\cap\mathcal{B}(o,D-r)^{c}\cap\mathcal{B}(o,r+D)}\exp(\lambda_2\mathcal{H}_1(r,\|y\|))\Bigg]\nonumber\\
&\stackrel{(c)}{=} \exp\left(-\lambda_2\pi r^2\right)\mathbb{E}_{\Phi_1}\Bigg[\prod_{{y\in\Phi_1\cap\mathcal{B}(o,D-r)}}\exp(\lambda_2\pi r^2)\Bigg]\times\nonumber\\
&\mathbb{E}_{\Phi_1}\Bigg[\prod_{y\in\Phi_1\cap\mathcal{B}(o,D-r)^{c}\cap\mathcal{B}(o,r+D)}\exp(\lambda_2\mathcal{H}_1(r,\|y\|))\Bigg] \nonumber\\
&\stackrel{(d)}{=} \exp\left(-\lambda_2\pi r^2\right)\exp\left(-\lambda_1\int_{\mathcal{B}(o,D-r)}(1-\exp(\lambda_2\pi r^2)){\rm d}y\right)\nonumber\\
&\exp\left(-\lambda_1\int_{\mathcal{B}(o,D-r)^{c}\cap\mathcal{B}(o,D+r)}(1-\exp(\lambda_2\mathcal{H}_1(r,\|y\|))){\rm d}y\right),
\end{align}
where $\Xi^{c}$ is the compliment of $\Xi$, step (c) is due to the fact that $\mathcal{B}(o,D-r)$ and $\mathcal{B}(o,D-r)^{c}\cap\mathcal{B}(o,r+D)$ are disjoint regions, and step (d) results from the direct application of the probability generating functional (PGFL) of the PPP~\cite{haenggi2012stochastic}. 
\begin{figure}
\centering
\includegraphics[width=0.5\columnwidth]{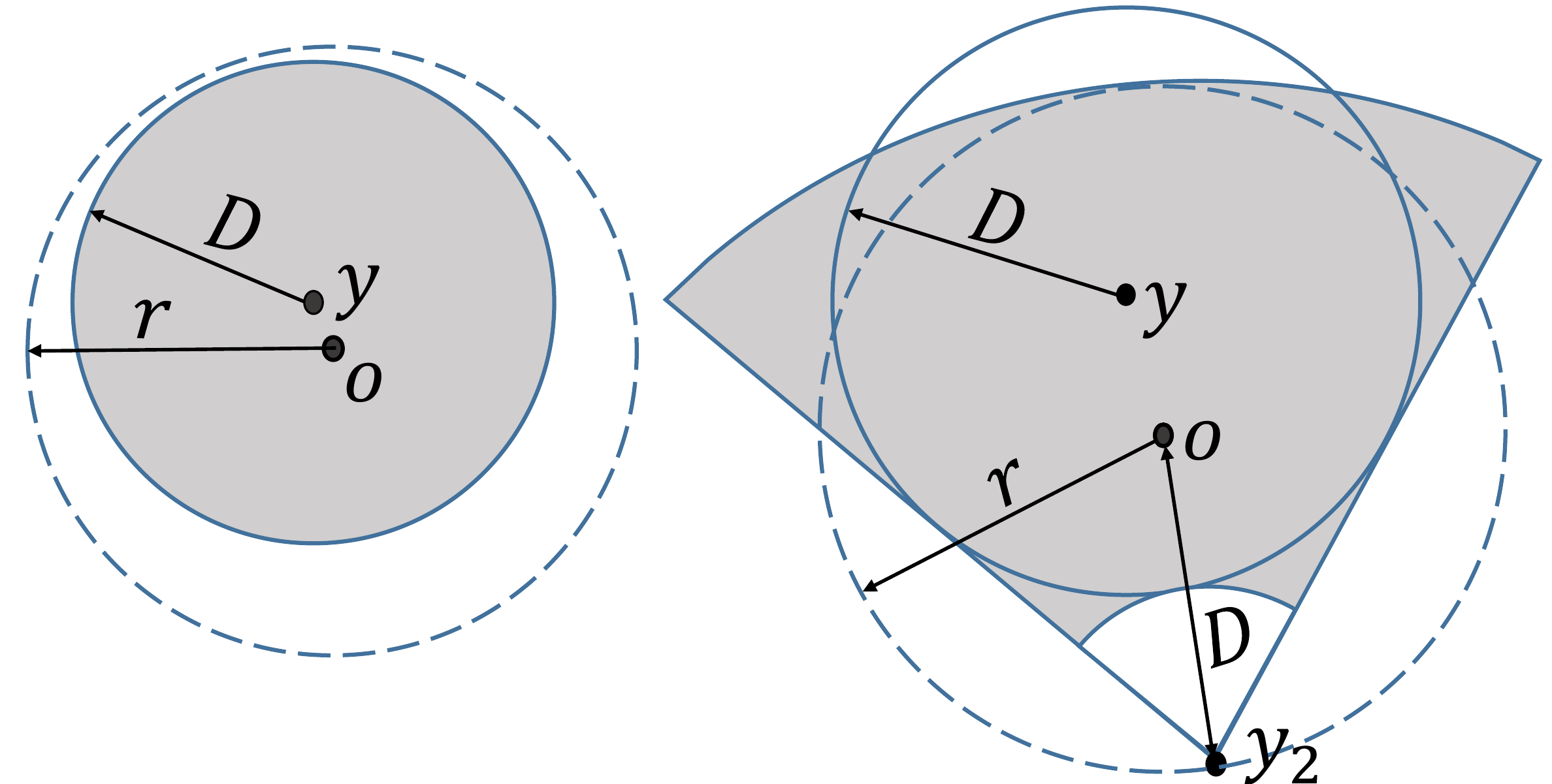}
\caption{The areas covered by $\tilde{A}(r)$ when the reference point is chosen uniformly at random from $\mathbb{R}^2$ independently of $\Psi$ ($r > D$).}
\label{fig:4}
\vspace{-5mm}
\end{figure}
For the case of $r>D$, we will follow the same approach as above. The area covered by $\tilde{\mathcal{A}}(r)$ is represented by two types of holes. The first type is when $\|y\|<r-D$. In that case, as shown in Fig.~\ref{fig:4} (left), the hole is completely enclosed inside the circle $\mathcal{B}(o,r)$. The second type is when $r-D<\|y\|<r+D$. In that case, as shown in Fig.~\ref{fig:4} (right), we model the intersection between the hole and the circle $\mathcal{B}(o,r)$ using the virtual point $y_2$ and the tangent lines as explained earlier. Following the same steps as above in the case of $r\leq D$ we get an upper bound on the CCDF as follows
\begin{align}
\label{11}
&\bar{F}_{R_1}(r)\leq \exp\left(-\lambda_2\pi r^2\right)\times\nonumber\\
&\exp\left(-\lambda_1\int_{\mathcal{B}(o,r-D)}(1-\exp(\lambda_2\pi D^2)){\rm d}y\right)\times\nonumber\\
&\exp\left(-\lambda_1\int_{\mathcal{B}(o,r-D)^{c}\cap\mathcal{B}(o,D+r)}(1-\exp(\lambda_2\mathcal{H}_1(r,\|y\|))){\rm d}y\right).
\end{align}
The final result follows from simple algebraic manipulations of the expressions in Eqs.~\ref{10} and ~\ref{11}.  

\subsection{Proof of Theorem~\ref{thm:hole_center}}\label{app:hole_center}
\begin{figure}
\centering
\includegraphics[width=0.7\columnwidth]{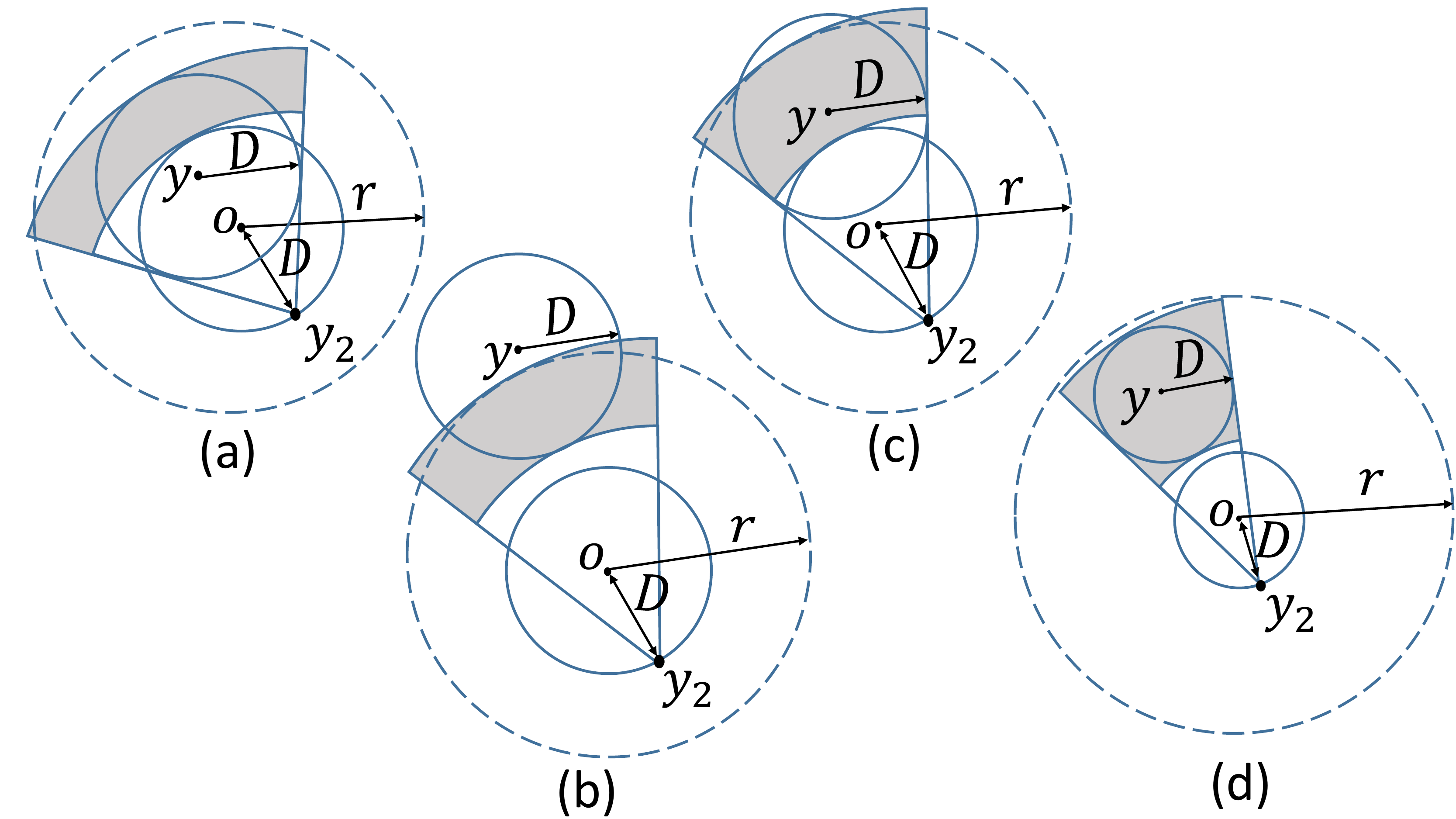}
\caption{The areas covered by $\tilde{A}(r)$ when the reference point is a hole center.}
\label{fig:5}
\vspace{-4mm}
\end{figure}
Since the reference point belongs to $\Phi_1$, the circle $\mathcal{B}(o,D)$ does not contain any points of the PHP. In other words, the minimum value of $R_2$ is $D$. When $r>D$, following the same approach as in Appendix~\ref{app:random_point} to upper bound the CCDF, the area covered by holes outside $\mathcal{B}(o,D)$ is upper bounded by $\tilde{A}(r)=\mathcal{B}(o,D)^{c}\cap\mathcal{B}(o,r)\cap\tilde{A}_1$ and $|\tilde{A}_1|=\sum_{y\in\Phi_1}|\mathcal{B}(y,D)|$. Hence, the CCDF is: $\bar{F}_{R_2}(r)\leq$ 
\begin{align}
&\mathbb{E}_{\Phi_1}\left[\mathbb{P}\left(\mathcal{N}_{\Phi_2}\left(\mathcal{B}(o,r)\setminus\{\mathcal{B}(o,D)\cup\tilde{\mathcal{A}}(r)\}\right)=0\bigg|\Phi_1\right)\right]\nonumber\\
&= \exp\left(-\lambda_2\pi (r^2-D^2)\right)\mathbb{E}_{\Phi_1}\left[\exp\left(\lambda_2|\tilde{\mathcal{A}}(r)|\right)\right].
\end{align}
The area covered by $\tilde{\mathcal{A}}(r)$ is represented by four types of holes. The first type is when $\|y\|\leq 2D,\|y\|\leq r-D$. The area in that case is as shown in Fig.~\ref{fig:5}.a. The second type, shown in Fig.~\ref{fig:5}.b, is when $\|y\|>2D,r-D<\|y\|\leq r+D$. The third type, shown in Fig.~\ref{fig:5}.c, is when $\|y\|<2D,r-D<\|y\|\leq r+D$. The fourth type, shown in Fig.~\ref{fig:5}.d, is when $\|y\|>2D,\|y\|\leq r-D$. The areas of the shaded sectors in the four cases can be combined in one mathematical expression,  which is $\mathcal{H}_2(r,r_y)$ as defined in Theorem~\ref{thm:hole_center}. Following similar procedure as in Appendix~\ref{app:random_point} leads to the final expression.
\vspace{-3mm}
\bibliographystyle{IEEEtran}
\bibliography{Dhillon_WCL2017-0337.bbl}

\end{document}

%% file: notation.tex
\def\nb0{{\mathbf{0}}}
\def\nb1{{\mathbf{1}}}

\def\nbbR{{\mathbb{R}}}

\newtheorem{definition}{Definition}

\newtheorem{theorem}{Theorem}

\newtheorem{cor}{Corollary}

